\documentclass[apj]{emulateapj}

\newcommand{\etal}{{et al.~}}
\newcommand{\msunh}{\>h^{-1}\rm M_\odot}
\newcommand{\Msun}{\>{\rm M_\odot}}

\newcommand{\rmd}{{\rm d}}
\newcommand{\Mh}{M_{\rm h}}
\newcommand{\Mq}{M_{\rm q}}
\newcommand{\Ms}{m_{\rm s}}
\newcommand{\Mstar}{M_{\ast}}
\newcommand{\calPc}{{\cal P_{\rm c}}}
\newcommand{\calPs}{{\cal P_{\rm s}}}
\newcommand{\Phis}{{\Phi_{\rm s}}}
\newcommand{\kmsmpc}{\>{\rm km}\,{\rm s}^{-1}\,{\rm Mpc}^{-1}}

\def\gtsima{$\; \buildrel > \over \sim \;$}
\def\ltsima{$\; \buildrel < \over \sim \;$}
\def\gta{\lower.7ex\hbox{\gtsima}}
\def\lta{\lower.7ex\hbox{\ltsima}}

\shorttitle{Subhalo - satellite connection}
\shortauthors{Yang et al.}

\begin{document}
            

\title{The subhalo - satellite connection and the fate of disrupted satellite
  galaxies}

\author{Xiaohu Yang\altaffilmark{1,4}, H.J. Mo \altaffilmark{2}, Frank C. van
  den Bosch\altaffilmark{3}}
\altaffiltext{1}{Shanghai Astronomical Observatory, the Partner Group of MPA,
  Nandan Road 80, Shanghai 200030, China; E-mail: xhyang@shao.ac.cn}
\altaffiltext{2}{Department of Astronomy, University of Massachusetts, Amherst
  MA 01003-9305}
\altaffiltext{3} {Max-Planck-Institute for Astronomy, K\"onigstuhl 17, D-69117
  Heidelberg, Germany}
\altaffiltext{4}{Joint Institute for Galaxy and Cosmology (JOINGC) of Shanghai
  Astronomical Observatory and University of Science and Technology of China}


\begin{abstract} In the standard  paradigm, satellite galaxies are 
  believed  to  be  associated  with  the population  of  dark  matter
  subhalos.   The assumption  usually  made is  that the  relationship
  between satellite galaxies and  sub-halos is similar to that between
  central  galaxies  and  host  halos.   In this  paper,  we  use  the
  conditional  stellar  mass  functions  of {\it  satellite  galaxies}
  obtained from a large galaxy group catalogue together with models of
  the  subhalo mass  functions  to explore  the  consequences of  such
  assumption in  connection to the stellar mass  function of satellite
  galaxies and the fraction and fate of stripped stars from satellites
  in galaxy groups and clusters  of different masses.  The majority of
  the  stripped stars  in massive  halos are  predicted to  end  up as
  intra-cluster stars, and the  predicted amounts of the intra-cluster
  component as a function of  the velocity dispersion of galaxy system
  match  well the observational  results obtained  by Gonzalez  et al.
  (2007).   The fraction of  the mass  in the  stripped stars  to that
  remain bound  in the central  and satellite galaxies is  the highest
  ($\sim  40\%$  of the  total  stellar  mass)  in halos  with  masses
  $M_h\sim  10^{14}\msunh$.   If  all   these  stars  end  up  in  the
  intra-cluster component (Max), or  a maximum amount of these stars
  is accreted into the central galaxy  (Min), then the maximum 
  fraction of the total  stars in  the whole universe that is 
  in the diffused  intra-cluster component is $\sim 19\%$,  and 
  the minimum is $\sim  5\%$. In the case of `Max',  
  the stellar mass  of the intra-cluster component in massive  
  halos with $M_h \sim  10^{15} \msunh$ is  roughly 6 times
  as large  as that of the  central galaxy.  This  factor decreases to
  $\sim 2$, $1$ and $0.1$ in halos with $M_h \sim 10^{14}$, $10^{13}$,
  and  $10^{12}  \msunh$, respectively.   The  total  amount of  stars
  stripped  from satellite galaxies  is insufficient  to build  up the
  central galaxies in halos  with masses $\la 10^{12.5}\msunh$, and so
  the  quenching of  star formation  must occur  in halos  with higher
  masses.   In  semi-analytical models  and  simulations  that do  not
  resolve the diffused component, caution must be exercised when using
  the  observed  stellar   mass/luminosity  function  of  galaxies  to
  constrain the star-formation, feedback  and merger processes in dark
  matter halos.
\end{abstract}




\section{Introduction}
\label{sec_intro}

Recent years  have seen a dramatic  impetus to link  galaxies to their
dark  matter  halos.   In  particular,  the  development  of  powerful
statistical  tools  such  as  the  halo  model,  the  halo  occupation
distribution  (HOD)  and the  conditional  luminosity function  (CLF),
combined with the  availability of large redshift surveys  such as the
two-Degree Field  Galaxy Redshift Survey (2dFGRS;  Colless \etal 2001)
and  the  Sloan Digital  Sky  Survey  (SDSS;  York \etal  2000),  have
resulted  in  reliable descriptions  of  how  galaxies with  different
properties are distributed over  halos of different masses (e.g.  Jing
\etal  1998; Peacock  \& Smith  2000; Berlind  \& Weinberg  2002; Yang
\etal 2003; van  den Bosch \etal 2003, 2007;  Zheng \etal 2005; Tinker
\etal 2005; Cooray 2006; Brown \etal 2008; Cacciato \etal 2008).

An  important aspect  of this  galaxy-dark matter  connection  is that
there  are two different  kinds of  galaxies: central  galaxies, which
reside at rest at the center  of their dark matter halo, and satellite
galaxies, which orbit  the halo associated with a  central galaxy.  It
is  generally believed  that satellite  galaxies themselves  reside in
dark matter subhalos.  Prior to being accreted into their current host
halo, satellite  galaxies were central galaxies,  and their associated
subhalos were host halos (throughout  this paper we use the term `host
halo' to  refer to a dark matter  halo which is not  a subhalo).  This
implies a direct link between the occupation statistics of dark matter
halos, and their merger/accretion histories.

While orbiting the host  halo, subhalos and their associated satellite
galaxies  are  subjected  to   dynamical  friction  which  causes  the
substructure to loose  its momentum and to sink  towards the center of
the  host halo. In  the mean  time, the  substructure is  subjected to
tidal forces which  try to dissolve it.  In  particular, tidal heating
and stripping cause  the system to loose mass, and  may even result in
the complete disruption of a  subhalo and its satellite galaxy.  There
are thus three possible fates for the stars in satellite galaxies: (i)
they  remain bound  to a  surviving  satellite galaxy,  (ii) they  are
accreted by  the central galaxy, or  (iii) they are  stripped from the
satellite galaxy, giving rise to  a stellar halo, which are stars that
orbit  the host halo  but that  are not  gravitationally bound  to any
particular galaxy.  Throughout this paper  we will refer to  the stars
that  belong  to  (ii)   and  (iii)  combined  as  the  `non-surviving
population'.  Obviously, the  satellites that are disrupted contribute
their entire  stellar content to  the central galaxy or  stellar halo,
while   the   survived   satellite   galaxies  may   also   contribute
significantly due to tidal stripping.

 The idea that satellite galaxies are stripped and/or disrupted is
  a   standard  prediction   of  hierarchical   models   of  structure
  formation. Numerous studies have addressed how this phenomenon gives
  rise to stellar  halos in systems ranging from  spiral galaxies like
  our own  Milky Way (e.g.  Searle  \& Zinn 1978;  Jonston \etal 1996;
  2001; Robertson \etal 2005; Font \etal 2006) to rich galaxy clusters
  (e.g. Gallagher \& Ostriker  1972; Merritt 1983; Mihos 2004; Willman
  \etal 2004; Lin \& Mohr 2004; Conroy \etal 2007; Purcell \etal 2007;
  2008;  Henriques  \etal 2008).  There  is  also ample  observational
  support for the  existence of stellar halos formed  out of disrupted
  satellite galaxies.   In particular, in  recent years it  has become
  clear that the stellar halo of  the Milky Way reveals a large amount
  of substructure  in the form  of stellar streams (Helmi  \etal 1999;
  Yanny \etal 2003; Bell \etal  2008), In some cases these streams can
  be  unambiguously associated with  their original  stellar structure
  (Ibata,  Gilmore \&  Irwin 1994;  Odenkirchen \etal  2002).  Similar
  streams have  also been  detected in our  neighbor galaxy  M31 (e.g.
  Ferguson  \etal 2002).   Unfortunately, due  to their  extremely low
  surface brightnesses,  it is very difficult to  detect stellar halos
  in   more   distant  galaxies   (but   see   Zibetti  \etal   2004).
  Consequently,  our  knowledge  of  stellar halos  around  individual
  galaxies  is  fairly  limited.   However,  groups  and  clusters  of
  galaxies also contain a significant stellar halo component, which is
  usually      referred      to      as     `intra-cluster      stars'
  (ICS)\footnote{Throughout this paper we will use the terms `ICS' and
    `stellar  halo' without  distinction.}.  In  fact,  data indicates
  that the  fraction of  stars associated with  such an  ICS component
  increases with increasing halo mass (Gonzalez, Zabludoff \& Zaritsky
  2007): in  massive clusters the mass  of the stellar halo  can be as
  large  as ten  times the  stellar  mass of  the central  (brightest)
  cluster  galaxy (Gonzalez,  Zabludoff  \& Zaritsky  2005; Seigar  et
  al. 2007).
  
With the use of high-resolution numerical simulations, it has recently
become possible to accurately determine the properties (mass function,
spatial  distribution,   orbits,  density  profiles,   spins)  of  the
population of  dark matter  subhalos (e.g.  Gao  \etal 2004;  De Lucia
\etal 2004; Tormen \etal 2004;  van den Bosch \etal 2005; Weller \etal
2005; Diemand \etal 2007; Giocoli \etal 2008).  If subhalos are indeed
associated with satellite galaxies, these properties should be related
to  the   luminosity  function,  spatial   distribution,  orbits,  and
structural properties of satellite galaxies.  Unfortunately, the exact
link  between  satellite galaxies  and  dark  matter  subhalos is  not
trivial.  Since the stellar component of a satellite is expected to be
more  tightly   bound  than  its  surrounding  dark   matter  (due  to
dissipation  during the formation  process), the  dark matter  is more
easily stripped, thus causing the ratio between stellar mass and total
mass to increase  with time.  Consequently, it is  to be expected that
the occupation statistics of subhalos as a function of their (current)
mass are different  from those of host halos.  However, since subhalos
were  host halos  before  being  accreted, it  seems  likely that  the
occupation statistics of subhalos as  a function of their mass {\it at
  the time of accretion} are identical  to those of host halos at that
time.  Indeed, it has been shown  that models based on this ansatz are
extremely  successful  in  explaining  the correlation  functions  and
luminosity functions of galaxies at different redshifts (e.g.  Vale \&
Ostriker  2004, 2006;  Conroy, Wechsler  \& Kravtsov  2006).  However,
since satellite galaxies  only make up a small  fraction ($\lta 30\%$)
of the total galaxy population  (e.g.  van den Bosch \etal 2007, 2008;
Tinker  \etal 2007;  Yang  \etal 2008a),  this  consistency cannot  be
considered  a sensitive test  of the  subhalo -  satellite connection.
This is  also apparent from the  fact that models  that link satellite
properties to  the {\it  current} subhalo mass  can also fit  the data
remarkably well (see e.g., Mandelbaum \etal 2006; Kim \etal 2008).

In  this  paper, we  use  the  conditional  stellar mass  function  of
satellite galaxies and the relation between stellar mass and halo mass
for  central  galaxies,  together  with  the  subhalo  mass  functions
obtained from recent numerical  simulations, to determine the relation
between satellite galaxies and dark matter subhalos. In particular, we
investigate  what  fraction   of  satellite  galaxies  survives,  what
fraction is  accreted into  the central galaxy,  and what  fraction is
tidally  disrupted.  Our  approach is  empirical, because  the stellar
mass functions and  the central stellar mass -  halo mass relation are
obtained from a  large SDSS galaxy group catalogue.   The structure of
this  paper is organized  as follows.  Section \ref{sec_data}  gives a
brief description  of the data and  the model used in  this paper.  In
Section~\ref{sec_disrupt}   we   present   our  predictions   of   the
conditional  stellar  mass   functions  for  satellite  galaxies.   In
Section~\ref{sec_mtq} we discuss the  possible fates of the stars that
are stripped  from the satellite galaxies.  Finally,  we summarize our
results in Section~\ref{sec_summary}.   Throughout this paper we adopt
a $\Lambda$CDM cosmology with  parameters that are consistent with the
three-year  data   release  of  the  WMAP   mission  (hereafter  WMAP3
cosmology):  $\Omega_{\rm  m} =  0.238$,  $\Omega_{\Lambda} =  0.762$,
$\Omega_{\rm  b} = 0.042$,  $n=0.951$, $h=H_0/(100  \kmsmpc)=0.73$ and
$\sigma_8=0.75$ (Spergel \etal 2007).

\section{Data and analysis}
\label{sec_data}

\begin{figure}
\plotone{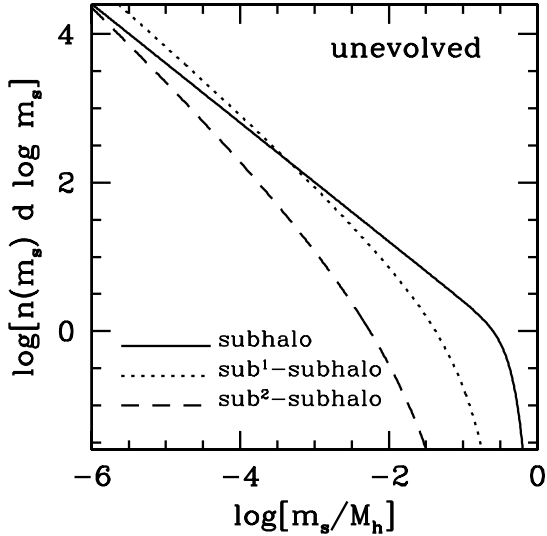}
\caption{The solid  line shows  the `unevolved' subhalo  mass function
  fraction   predicted  by   Giocoli  et   al.   (2008)   from  N-body
  simulations.   The dotted and  dashed lines  are the  predicted mass
  functions of sub$^1$-subhalo and sub$^2$-subhalos, respectively (see
  Eq.[~\ref{eq:subsubi}]).}
\label{fig:subsub}
\end{figure}
\begin{figure*}
\plotone{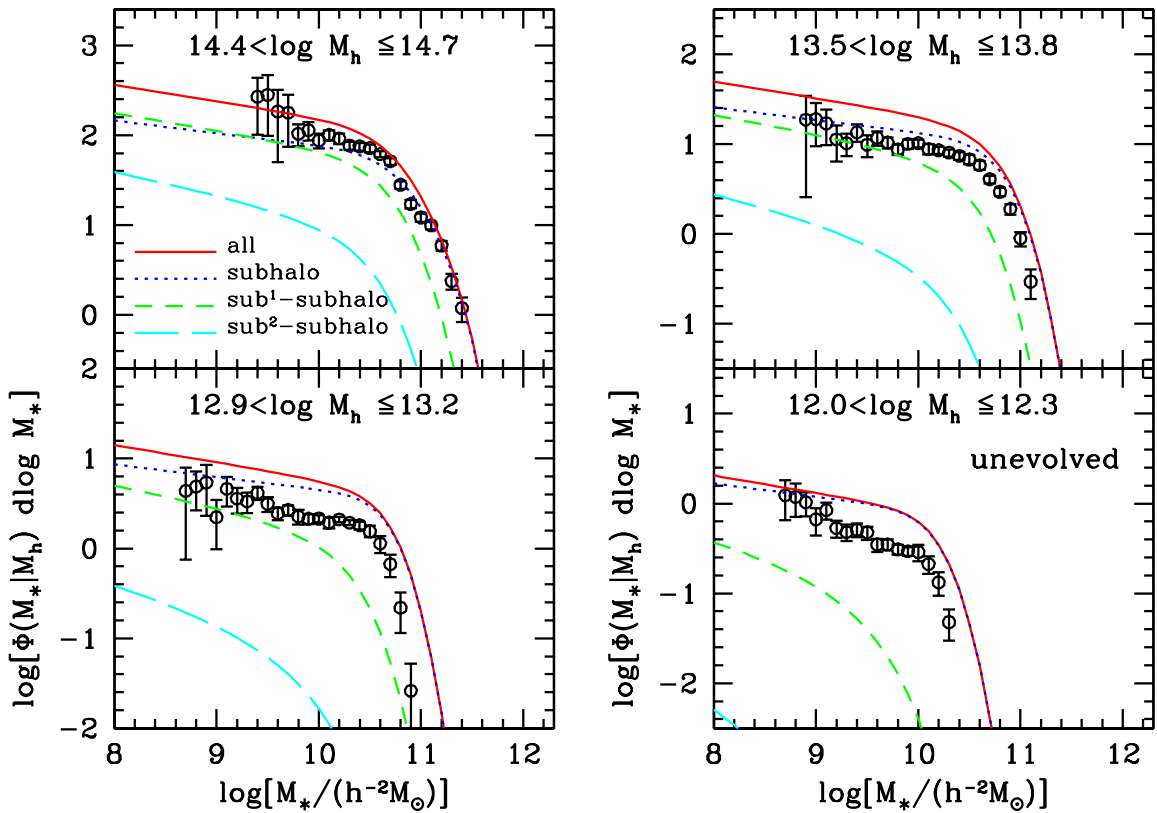}
\caption{The conditional stellar mass functions for satellite galaxies
  in  halos of  different  masses  as indicated  in  each panel.   The
  circles are the measurements from the SDSS DR4 group catalogue.  The
  error-bars  represent  scatter   among  four  different  samples  as
  described in the text.  Dotted, dashed and long-dashed lines in each
  panel  show  the  predicted   CSMFs  using  the  unevolved  subhalo,
  sub$^1$-subhalo  and sub$^2$-subhalo  mass  functions, respectively.
  Here we assume that the  relation between stellar mass and halo mass
  is the same for satellite  galaxies and central galaxies.  The solid
  lines are  the combinations of  the three components.  Note  that we
  did  not  take  into  account  the  contribution  to  the  satellite
  population, which  is negligible, from higher  order subhalos, e.g.,
  sub$^3$-subhalos and sub$^4$-subhalos etc.  }
\label{fig:data}
\end{figure*}
\begin{figure*}
\plotone{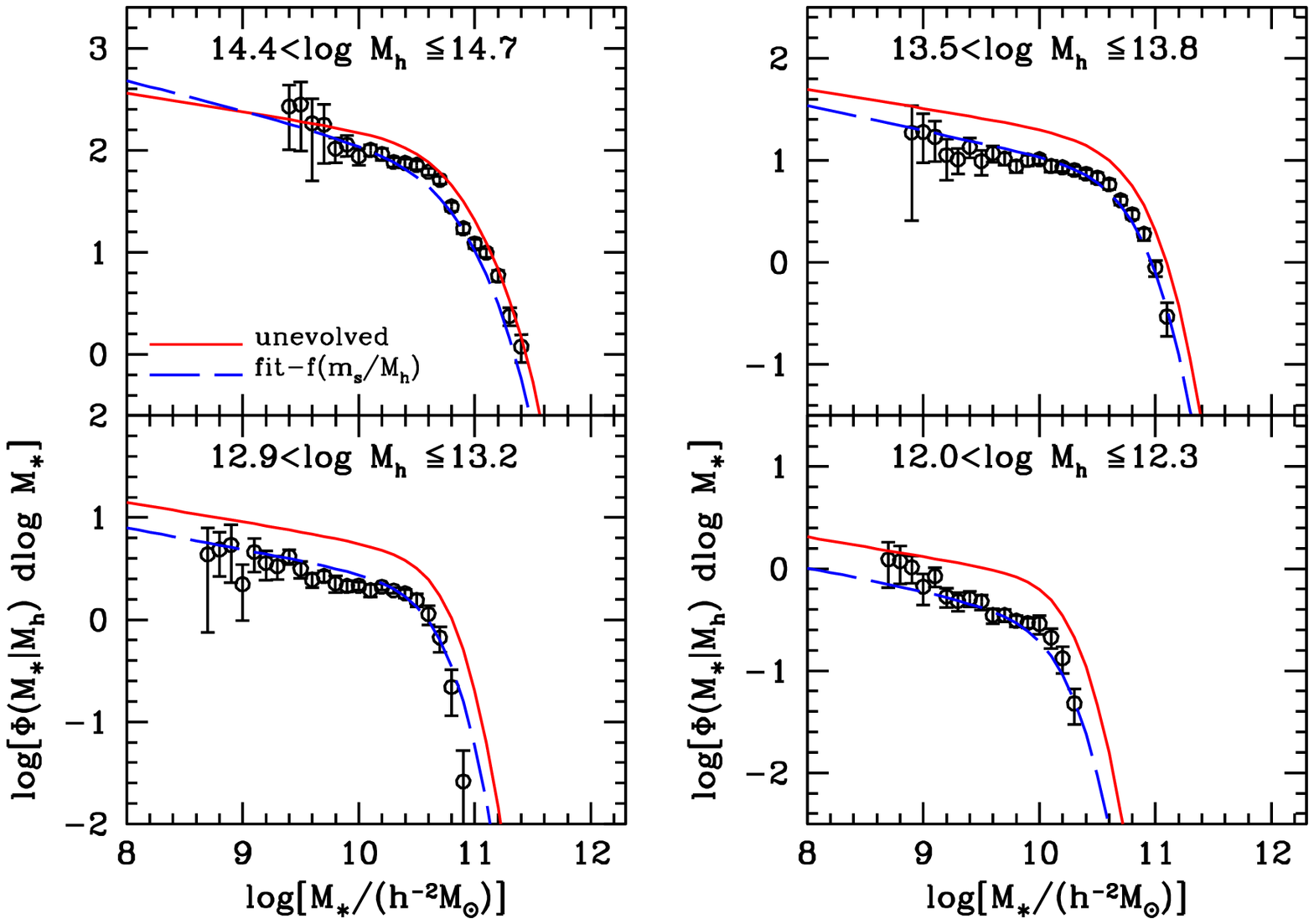}
\caption{Same as  Fig. \ref{fig:data}, but here we  show the predicted
  total CSMFs using the unevolved  (solid lines) SHMF and the best fit
  results  (long-dashed lines)  of the  model described  in  the text,
  respectively.}
\label{fig:evolved}
\end{figure*}

The  data  used in  this  paper  is based  on  the  SDSS galaxy  group
catalogues of Yang \etal  (2007; hereafter Y07).  These catalogues are
constructed by applying the  halo-based group finder developed by Yang
\etal (2005)  to the New York University  Value-Added Galaxy Catalogue
(NYU-VAGC;  see Blanton  \etal  2005),  which is  based  on SDSS  Data
Release 4  (Adelman-McCarthy \etal  2006).  Detailed tests  have shown
that  this group  finder is  very successful  in  associating galaxies
according to their common dark  matter halos, and that the halo masses
that have been assigned to the groups are reliable.

In  Yang  \etal (2008a,b)  we  have  used  these group  catalogues  to
determine   the  conditional  luminosity   functions  (CLF)   and  the
conditional stellar  mass functions (CSMF) separately  for central and
satellite galaxies.  Central galaxies  are defined as the most massive
group members in  terms of their stellar mass,  and satellites are the
group members that are not  centrals.  The CSMF of satellite galaxies,
$\Phis(\Mstar|\Mh)$,  gives  the  average  number of  satellites  with
stellar masses in  the range $\Mstar \pm \rmd\Mstar$  that reside in a
host halo of mass $\Mh$.  The open circles in Fig.~\ref{fig:data} show
the  $\Phis(\Mstar|\Mh)$  obtained  by  Yang \etal  (2008b,  hereafter
YMB08) for  halos in  four different mass  bins.  These CSMFs  are the
averages obtained from two group samples (Samples II and III) and from
using two different halo mass estimators (see YMB08 for details).  The
errorbars in Fig.~\ref{fig:data} reflect  the scatter among these four
results, which in general are  much larger than the statistical errors
obtained using bootstrap samples.

In order  to link  the satellites to  the subhalo population,  we also
need the subhalo mass function  (SHMF).  In fact, since it is expected
that the  properties of satellite galaxies  are linked to  the mass of
their subhalos  {\it at  their time of  accretion} (see  discussion in
\S\ref{sec_intro}), we need  the so-called ``unevolved'' SHMF, $n_{\rm
  un}(\Ms|\Mh)$,  which gives the  number of  subhalos that  have been
accreted by the  main progenitor of a halo of  present day mass $\Mh$,
as a  function of  their mass  $\Ms$ at the  time of  accretion.  This
unevolved  SHMF should  not  be confused  with  the ``evolved''  SHMF,
$n_{\rm ev}(\Ms'|\Mh)$,  which gives the number of  subhalos with {\it
  present-day} masses in the range $\Ms' \pm \rmd\Ms'$ that reside, at
present, in a host halo of mass $\Mh$. The unevolved SHMF evolves into
the evolved  SHMF due  to the combined  effect of  dynamical friction,
tidal  stripping and  tidal  heating, which  causes  some subhalos  to
dissolve, and others  to loose mass (see van den  Bosch \etal 2005 for
details).

Using  high-resolution  numerical  simulations, Giocoli  \etal  (2008)
found that the unevolved SHMF is accurately described by
\begin{equation}\label{eq:shmf0}
n_{{\rm un},0}(\Ms|\Mh) = {0.176\over\Mh} \,
\left({\Ms\over\Mh}\right)^{-1.8}
\exp\left[-12.27 \left({\Ms\over\Mh}\right)^{3}\right]\,.
\end{equation}
Note, however, that this SHMF only includes the population of subhalos
that are {\it directly} accreted  onto the main progenitor of the host
halo (hence  the subscript `0').   This neglects the  possibility that
subhalos may  themselves contain `sub-subhalos',  and sub-subhalos may
contain `sub-sub-subhalos'  and so on.   For convenience, we  will use
the  notation  `sub$^i$-subhalo'  ($i=1,2,3,4...$)  to  refer  to  the
$i^{\rm th}$ level of  subhalos.  In modeling the satellite population
associated  with   the  subhalos,  neglecting   the  `sub$^i$-subhalo'
populations may result in an  underestimate of the number of satellite
galaxies.   This is  especially true  for massive  systems,  where the
merging progenitors  may already contain  relatively massive satellite
galaxies.

Fortunately,  because of  the self-similar  behavior of  the unevolved
SHMF  (Eq.   \ref{eq:shmf0}),  we  can calculate  the  unevolved  SHMF
including    the   sub$^i$-subhalo    populations.     Assuming   that
Eq.~(\ref{eq:shmf0})   also   applies  to   subhalos,   the  SHMF   of
sub$^i$-subhalos can be written as
\begin{equation}\label{eq:subsubi} 
n_{{\rm  un},  i}(m_{{\rm  s},  i}|\Mh)  =
\int\limits_0^{\infty}  n_{{\rm  un},0}(m_{{\rm s},  i}|\Ms)  \, n_{{\rm  un},
i-1}(\Ms|\Mh) \,\rmd\Ms\,.
\end{equation}
Fig.~\ref{fig:subsub} shows the predictions for the unevolved subhalo,
sub$^1$-subhalo and sub$^2$-subhalo mass functions, thus obtained.  As
one   can   see,   the   contributions   from   sub$^1$-subhalos   and
sub$^2$-subhalos to  the total SHMF can  be larger than  that from the
subhalos at $\log (\Ms/\Mh) \la  -3$ and $\la -6$, respectively.  They
can  thus  contribute  a  significant fraction  of  (small)  satellite
galaxies.

The final  ingredient for our modeling is  the conditional probability
distribution, $\calPc(\Mstar|\Mh)$, that a  halo of mass $\Mh$ hosts a
{\it  central} galaxy  with  stellar mass  $\Mstar$.   Using our  SDSS
galaxy group catalogues, YMB08 found that $\calPc(\Mstar|\Mh)$ is well
described by a log-normal distribution whose median is given by
\begin{equation}\label{eq:Ms_fit}
\langle\Mstar\rangle(\Mh) = 
M_0 {(\Mh/M_1)^{\alpha +\beta} \over (1+\Mh/M_1)^\beta} \,.
\end{equation}
Here  $M_1$ is  a characteristic  halo  mass so  that $\Mstar  \propto
\Mh^{\alpha+\beta}$   for   $\Mh   \ll   M_1$  and   $\Mstar   \propto
\Mh^{\alpha}$ for $\Mh \gg M_1$. The parameters obtained from the SDSS
groups are:  $\log M_0= 10.306$, $\log M_1=  11.040$, $\alpha= 0.315$,
and $\beta= 4.543$, where $M_0$ is in units of $h^{-2}\Msun$ and $M_1$
in $\msunh$ (see YMB08 for details). The width of $\calPc(\Mstar|\Mh)$
is found to  be roughly independent of halo mass  (see also More \etal
2008), with a dispersion $\sigma(\log\Mstar) = 0.173$.

\section{The disruption of satellite galaxies}
\label{sec_disrupt}

The prediction of the CSMF of satellite galaxies can be written as
\begin{equation}\label{phismodel}
\Phis(\Mstar|\Mh) = \int_0^{\infty} \calPs(\Mstar|\Ms) \,
n_{\rm un}(\Ms|\Mh) \rmd\Ms\,,
\end{equation}
where $\calPs(\Mstar|\Ms)$  is the probability that a  subhalo of mass
$\Ms$ at  the time of  accretion hosts a present-day  satellite galaxy
with stellar mass $\Mstar$, and
\begin{equation}\label{totshmf}
n_{\rm un}(\Ms|\Mh) = \sum_{i=0}^{N_{\rm max}} n_{{\rm un},i}(\Ms|\Mh)
\end{equation}
is  the unevolved  SHMF, including  all sub$^i$-subhalos  up  to level
$N_{\rm max}$.  We start by assuming that the stellar mass - halo mass
relation does not evolve  with redshift, so that $\calPs(\Mstar|\Ms) =
\calPc  (\Mstar  | \Mh=\Ms)$.   In  this  case,  we denote  the  model
prediction of $\Phis$ by $\Phi_{\rm un}$, and so
\begin{equation}\label{phiunmodel}
\Phi_{\rm un}(\Mstar|\Mh) = \int_0^{\infty} 
\calPc(\Mstar|\Ms) \,n_{\rm un}(\Ms|\Mh) \rmd\Ms\,.
\end{equation}
The solid lines in each  of the panels of Fig.~\ref{fig:data} show the
CSMFs  thus  obtained with  Eq.~(\ref{phiunmodel})  and using  $N_{\rm
  max}=2$.   The  dotted,  dashed   and  long-dashed  lines  show  the
contributions due to  subhalos, sub$^1$-subhalos and sub$^2$-subhalos,
respectively,   obtained  by   replacing   $n_{\rm  un}(\Ms|\Mh)$   in
Eq.~(\ref{phiunmodel}) with $n_{{\rm un},i}(\Ms|\Mh)$ for $i=0, 1$ and
$2$.   Note that  for  massive halos  ($\Mh  \gta 10^{14}\msunh$)  the
contribution  of  sub$^1$-subhalos  is   comparable  to  that  of  the
subhalos, indicating  that their  contribution cannot be  ignored.  In
low mass halos ($\Mh \lta 10^{13}\msunh$), on the other hand, ignoring
the  sub$^i$-subhalos for  $i \geq  1$ makes  little  difference.  The
satellite  population contributed by  the sub$^2$-subhalos  is roughly
10\% in  massive halos with $\Mh  \sim 10^{14}\msunh$ but  drops to $<
1\%$  in halos  with  $\Mh  \sim 10^{12}\msunh$.   We  found that  the
overall  contribution by  sub$^i$-subhalos with  $i>2$ is  always less
than a few  percent and can be safely neglected.   In what follows, we
therefore  adopt   $N_{\rm  max}=2$  throughout,   and  unless  stated
otherwise, we refer to subhalos, sub$^1$-subhalos and sub$^2$-subhalos
collectively as `subhalos'.

\begin{figure}
\plotone{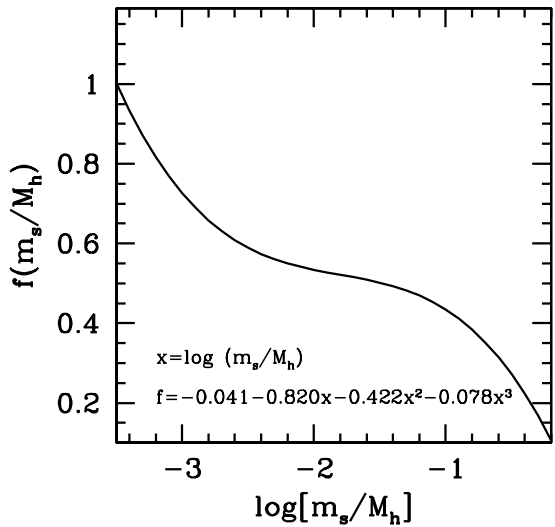}
\caption{The best-fit  (survival) fraction  of subhalos that  can host
  satellite  galaxies  in the  same  way  as  halos can  host  central
  galaxies at present time $z=0$.  Here we assume that the subhalos in
  host halos of different masses  have the same survival fraction as a
  function of  $\Ms/\Mh$.  As shown  in Fig.  \ref{fig:data}  the best
  fit predictions of this model are remarkably good.  }
\label{fig:f}
\end{figure}

Comparing   the  model   predictions  with   the   observational  data
(open-circles with  error-bars) shown in Fig.   \ref{fig:data}, we see
that the predicted CSMFs for satellite galaxies are roughly consistent
with the  data for massive  groups.  However, the  model significantly
{\it  over-predicts} the  CSMFs  for low-mass  halos.   There are  two
different possible explanations for this discrepancy.  First, halos of
a given mass  at high redshift may contain  different amounts of stars
from halos  at $z=0$ with  the same mass,  so that the stellar  mass -
halo mass relation for subhalos  is different from that of present-day
central galaxies.   In order to  explain the discrepancy,  the stellar
mass  fractions of halos  of a  given mass  then have  to be  lower at
higher redshifts.   Furthermore, since  the extent of  the discrepancy
depends on halo mass, the  amplitude of the redshift dependence has to
be different for halos of  different masses.  This solution is not
  very likely,  since it is rather  contrived to assume  that the star
  formation efficiency in progenitor halos  depends on the mass of the
  halo  in  which it  will  end up  in  the  future. Nevertheless,  we
  acknowledge that the average  relation between halo mass and stellar
  mass may  well evolve  with redshift (see  e.g., Conroy  \& Wechsler
  2008    for    empirical   constraints    in    support   of    such
  evolution). However, given the typical accretion times for subhalos,
  we believe that  this will not have a strong  impact on our results.
  This is also supported by the work of Purcell \etal (2007), who have
  shown that the  predictions for the stellar halo  mass fractions are
  extremely robust to  changes in the star formation  histories of the
  galaxies.   The  second,  more  likely,  possibility  is  that  the
subhalos  and their  satellite  galaxies experience  mass loss  and/or
disruption due to the combined  effect of dynamical friction and tidal
forces.

In this  paper we  focus on  the second possibility.  In this  case, a
satellite galaxy  can either be  completely disrupted, hence  does not
contribute to  the satellite population,  or experience mass  loss but
survives as  a satellite of {\it  lower} mass.  Both  of these effects
can change  the predicted stellar mass function  of satellite galaxies
relative to  $\Phi_{\rm un}$.  As  an illustration, let us  consider a
simple model  in which a  satellite is either completely  disrupted or
remains  intact.  This  assumption is  consistent with  the simulation
results  that a satellite  is quickly  disrupted after  it has  lost a
significant amount of  mass (Moore et al. 1999).  It  is also valid if
the  disruption of  a  satellite is  due  to merger  into the  central
galaxy. As shown  in van den Bosch (2005) and  Giocoli et al.  (2008),
on average  the instantaneous mass loss  rate of a  subhalo depends on
the ratio  between the  instantaneous subhalo mass  and the  host halo
mass at the time in question. Here we ignore such details.  Instead we
consider  a  simple model  where  the  survivor  fraction, $f$,  is  a
function of  the ratio between  the subhalo mass at  accretion, $\Ms$,
and the host  halo mass at the present time,  $\Mh$.  Motivated by the
roughly  self-similar behavior  of the  subhalo population,  we assume
that $f(\Ms/\Mh)$  is universal in  halos of different masses.   For a
given  $f(\Ms/\Mh)$ the  CSMF is  given by  Eq.~(\ref{phismodel}), but
with  $n_{\rm  un}(\Ms|\Mh)$  replaced   by  $  f(\Ms/\Mh)  \,  n_{\rm
  un}(\Ms|\Mh)$.  We model  $f(x)$ using a polynomial form,  $f(x) = a
+bx  +cx^2 +dx^3$, and  determine the  free parameters  $(a,b,c,d)$ by
fitting the  model to nine CSMFs  that cover the mass  range $12.0 \le
\log \Mh \le 14.7$,  each with a 0.3 dex bin width  in halo mass.  The
CSMFs corresponding to the best-fit model are shown as the long-dashed
lines in Fig.~\ref{fig:evolved}.  Note that this simple model fits the
data surprisingly  well, supporting  the assumption that  the survivor
fraction  depends  only on  $\Ms$  and  $\Mh$  via their  ratio.   The
best-fit $f(\Ms/\Mh)$ is shown  in Fig.\,\ref{fig:f}, which shows that
the survivor fraction decreases  monotonically with subhalo mass, from
$\sim 1$ for  subhalos with $\Ms \sim 10^{-3.5}\Mh$  to $\sim 0.1$ for
subhalos  with $\Ms  \sim 0.6\Mh$  (recall  that $\Ms$  refers to  the
subhalo mass  at the time of  accretion). Thus, if the  stellar mass -
halo mass  relation does not  evolve with redshift, then  more massive
(relative  to  their host)  subhalos  and  their associated  satellite
galaxies are predicted to have a smaller survival probability. This is
consistent with  a picture in which dynamical  friction is responsible
for transporting satellite galaxies to the inner regions of their host
halos, where they  are more likely to be disrupted  by tidal forces or
to merge with the central galaxy.

\begin{figure*}
\plotone{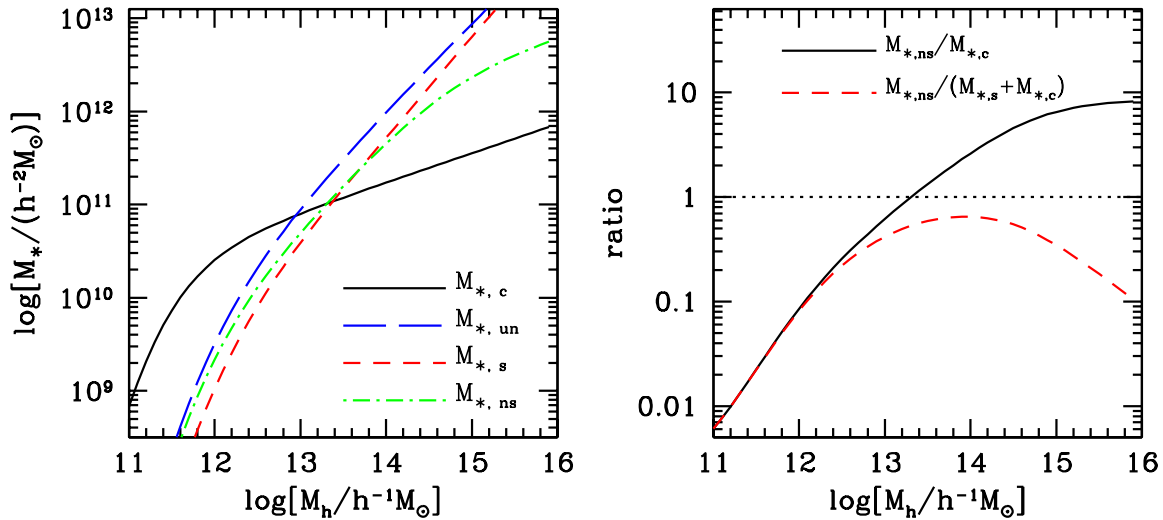}
\caption{Left panel: the long-dashed  line shows the predicted average
  total  stellar  mass  that  the  satellite  galaxies  (in  unevolved
  subhalos) can bring into the  host halos.  The dashed line shows the
  average total  stellar mass remain  in the satellite  galaxies which
  can  be observed  today.   The dot-dashed  line  corresponds to  the
  average total stellar mass that  has been stripped or disrupted from
  the satellite  galaxies.  For comparison,  the solid line  shows the
  predicted average  stellar mass of the central  galaxy. Right panel:
  the ratios of the stripped  or disrupted stellar mass from satellite
  galaxies, $M_{\rm \ast, ns}$, to that of the central galaxy, $M_{\rm
    \ast, c}$, (solid  line) and to the total  stellar mass in central
  and satellite  galaxies, $M_{\rm \ast, c}+M_{\rm  \ast, s}$, (dashed
  line).  }
\label{fig:MM}
\end{figure*}

\section{Merging, Tidal Disruption and Quenching}
\label{sec_mtq}

We  can use  the stellar  mass  - halo  mass relation  to predict  the
following stellar mass components for halos of a given mass: the total
stellar mass  that the  satellite galaxies bring  into the  host halo,
which is given by
\begin{equation}
M_{\rm  \ast, un}(\Mh) = \int_0^{\infty} \rmd\Mstar \, \Mstar
\Phi_{\rm un}(\Mstar\vert \Mh)\,,
\end{equation}
and the total stellar  mass in surviving satellite galaxies,
\begin{equation}
M_{\rm  \ast, s}(\Mh) = \int_0^{\infty} \rmd\Mstar \, \Mstar
\Phis (\Mstar\vert \Mh)\,.
\end{equation}
Note that both $M_{\rm \ast,  un}(\Mh)$ and $M_{\rm \ast, s}(\Mh)$ can
be  obtained  without  assuming  whether  a  satellite  is  completely
disrupted or  only partly stripped.  The difference  between these two
stellar masses,  $M_{\rm \ast,  ns} (\Mh) =  M_{\rm \ast, un}  (\Mh) -
M_{\rm  \ast,   s}  (\Mh)$,   is  the  total   stellar  mass   of  the
non-survivors, which consists of  the stars in satellite galaxies that
are  completely  disrupted  and  those  that  are  stripped  from  the
satellite galaxies.  In other words,  $M_{\rm \ast, ns}$ is the sum of
the stellar mass of the stellar  halo (i.e., the ICS) plus the stellar
mass  of the  central galaxy  that  has been  accreted from  satellite
galaxies.

The left-hand  panel of Fig.   \ref{fig:MM} shows the  predictions for
$M_{\rm \ast, un}$ (long-dashed line), $M_{\rm \ast, s}$ (short-dashed
line), and $M_{\rm  \ast, ns}$ (dot-dashed line), all  as functions of
halo mass.  For  comparison, the solid line shows  the stellar mass of
central galaxies  as a function of  halo mass [i.e.   the stellar mass
  -halo mass relation given by Eq.\,(\ref{eq:Ms_fit})].  Note that, in
massive  halos with $\Mh\ga  10^{13}\msunh$, the  stellar mass  in the
non-surviving component is larger than the stellar mass of the central
galaxy.

The dashed  line in the  right-hand panel of  Fig.\,\ref{fig:MM} shows
$M_{\rm  \ast, ns}/  (M_{\rm \ast,c}  + M_{\rm  \ast, s})$,  the ratio
between  the  stellar mass  of  the  non-surviving  component and  the
combined  stellar  mass  of  the  central  galaxy  and  the  surviving
satellites.  In terms  of $M_{\rm \ast, ns}/ (M_{\rm  \ast,c} + M_{\rm
  \ast,  s})$, halos with  $\Mh \sim  10^{14}\msunh$ have  the highest
ratio, $\sim  0.6$, i.e. $\sim 40\%$  of the total stellar  mass is in
the non-surviving  component.  In less massive halos,  the ratio drops
rapidly due to the fact that  the stellar mass - halo mass relation is
extremely steep at the low mass end, so that low mass subhalos contain
very few  stars even at their  time of accretion.  At  the massive end
the ratio  also decreases due to  two effects. First of  all, for $\Mh
\gta  10^{12}   \msunh$  the  stellar  mass   fractions  decline  with
increasing halo mass (i.e., the slope  of the stellar mass - halo mass
relation is less than unity), so that more massive subhalos contribute
fewer stars per unit dark  matter mass.  Secondly, many of the fainter
satellite galaxies, which reside in less massive subhalos, can survive
since  dynamical friction  is not  effective for  subhalos  with small
$\Ms/\Mh$.

The  ratio of stars  that are  in the  non-surviving component  in the
entire  Universe to  those that  are locked  up in  either  central or
satellite galaxies can be estimated using:
\begin{equation}
R_{\rm ns}\equiv 
{\int_0^{\infty}  M_{\rm \ast, ns}(\Mh) n(\Mh) d\Mh
\over 
\int_0^{\infty}\left[M_{\rm \ast,c}(\Mh) 
+ M_{\rm \ast, s}(\Mh)\right] n(\Mh) d\Mh}\,,
\end{equation} 
where $n(\Mh)$  is the halo mass  function (e.g., Sheth,  Mo \& Tormen
2001; Warren et al.  2006).  We find that $R_{\rm ns} \sim 23 \%$.  If
we assume  that all  of the stars  in the non-surviving  component are
turned into ICS,  then $\sim 19 \%$, which is  the upper-limit, of the
stellar  mass in the  universe is  in ICS.  On the  other hand,  if we
assume  that  maximum of  these  stars  are  accreted by  the  central
galaxies, then $\sim  5 \%$, which is the  lower-limit, of the stellar
mass in the universe is in ICS.

\subsection{The fate of stripped and dispersed stars}
\label{sec:fate}

We  now investigate  the  fate of  the  stars that  are stripped  from
satellite  galaxies.   In  particular,  we examine  what  fraction  is
actually disrupted by tidal forces, thus giving rise to a stellar halo
(i.e.  the  ICS),  and  what  fraction  is  accreted  by  the  central
galaxy. For this purpose, we show  as the solid line in the right-hand
panel of Fig.\,\ref{fig:MM}, the ratio between the stellar mass of the
non-surviving component,  $M_{\rm \ast, ns}$, and that  of the central
galaxy $M_{\rm  \ast, c}$.  As one  can see, in small  mass halos with
$\Mh  \la 10^{13}\msunh$,  the  mass  of the  stripped  stars is  much
smaller than  that of the  central galaxy, indicating that  the latter
cannot have  grown substantially due  to the accretion  of satellites.
Rather,  central   galaxies  in  low   mass  halos  must   have  grown
predominantly via star formation.  It  also means that if all the mass
of the  stripped stars  ends up as  a stellar  halo, the mass  of that
stellar halo can  only be a small fraction of the  mass of the central
galaxy.  In  halos with  $\Mh\ga 10^{13} \msunh$,  on the  other hand,
$M_{\rm \ast, ns} \gg M_{\rm \ast,c}$, and a very significant fraction
of  the stellar mass  of the  central galaxy  may consist  of accreted
stars stripped  from satellites  (but does not  have to).  It  is also
clear  that we  can exclude  the possibility  that {\it  all} stripped
stars  are accreted  into  the  central galaxy,  as  this would  imply
stellar masses  for the  brightest cluster galaxies  that are  about 8
times higher than observed. Rather, the stripped stars must have given
rise to a substantial stellar halo.

\begin{figure}
\plotone{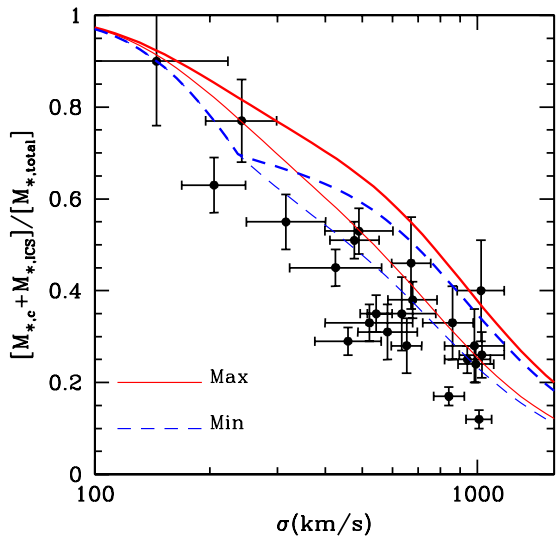}
\caption{ The fraction of the total stellar masses, which is contained
  in  the central  galaxy and  intra-cluster  stars as  a function  of
  cluster velocity  dispersion. Data with errorbars  are obtained from
  Gonzalez et al.  (2007), measured within $r_{500}$ for the 23 groups
  and clusters  in their sample.  The  solid and dashed  lines are our
  model predictions for the two extreme cases. See text for details.}
\label{fig:ICS}
\end{figure}

The solid dots with  errorbars in Fig.~\ref{fig:ICS} show the observed
fractions of  the total  stellar mass present  in groups  and clusters
that is contained  in the central galaxy and the ICS  as a function of
the line-of-sight  velocity dispersion of the  member galaxies.  These
data are taken from Gonzalez,  Zaritsky \& Zabludoff (2007), under the
assumption that  satellite galaxies, central galaxies and  the ICS all
have the same mass-to-light ratios  (in the $i$-band).  We can compare
the data with two extreme predictions of our model, by converting halo
mass into  a line-of-sight velocity  dispersion using equation  (6) in
Yang \etal (2007).  The solid lines, labeled `Max', corresponds to the
prediction  under   the  assumption  that  {\it  all}   stars  of  the
non-surviving component are added to  the stellar halo (i.e., no stars
stripped from satellite galaxies  are accreted by the central galaxy).
Clearly, this corresponds to the  maximum amount of ICS possible.  The
dashed lines, labeled `Min', corresponds to the minimum amount of ICS,
i.e.  the stellar mass in the  ICS is assumed to be $\max(M_{\rm \ast,
  ns}-M_{\rm  \ast, c},  0)$.  Note  that the  data of  Gonzalez \etal
(2007)  are obtained  within $r_{500}$,  the radius  within  which the
cluster mass  density exceeds the critical  value by a  factor of 500.
In our model  prediction, the total mass of  satellite galaxies within
$r_{500}$ is estimated by  assuming that the distribution of satellite
galaxies follow  the NFW (Navarro,  Frenk \& White 1997)  profile with
concentration  appropriate for  the  halo mass  in  question. For  the
distribution of  ICS, we consider two  cases. Case I  assumes that the
ICS  have the  same distribution  as the  satellite galaxies,  and the
corresponding results are shown  in Fig.~\ref{fig:ICS} as the two thin
lines. Case II assumes that  all ICS are distributed within $r_{500}$,
and the  corresponding results are shown in  Fig.~\ref{fig:ICS} by the
two thick lines.  It is reassuring that, for a given assumption of the
ICS distribution, the two extreme  models, `Max' and `Min', give quite
similar results.  The prediction of Case  I is in  good agreement with
the  data  over the  entire  range of  masses  probed,  while Case  II
over-predicts  the  ratio  moderately.   It is  unclear  whether  this
moderate discrepancy necessarily implies  that the distribution of the
ICS extends beyond $r_{500}$, because there are other factors that can
cause such discrepancy.  For example, stars in the  central galaxy and
in  the  ICS  component  may,   on  average,  have  a  higher  stellar
mass-to-light ratio  than those in  the surviving satellites,  so that
the ratio $(M_{\rm \ast,  c}+M_{\rm\ast, ICS})/M_{\rm \ast, total}$ is
underestimated in the  data due to the assumption  of the same stellar
mass-to-light ratio  for all components.  In addition,  there are also
uncertainties in the estimates  of the velocity dispersions, $\sigma$,
particularly for poor systems.  Given these uncertainties, we consider
the  overall  agreement between  the  data  and  the model  prediction
remarkable.  Clearly, better observational  data are required in order
to distinguish the different models considered here.

The  idea that  a  significant  fraction of  the  stars stripped  from
satellite  galaxies end up  as ICS  is not  only consistent  (in fact,
required)  by the  data, but  has  also been  found in  hydrodynamical
simulations of galaxy clusters (e.g. Napolitano \etal 2003; Murante et
al. 2004, 2007; Willman  \etal 2004; Sommer-Larsen, Romeo \& Portinari
2005; Rudnick, Mihos \& McBride 2006). In addition, as shown by Monaco
\etal (2006)  and Conroy \etal  (2007), the creation of  a significant
ICS component is  also required in order to reconcile  the low rate at
which  massive galaxies have  grown since  $z=1$, with  the relatively
high rate at which their host halos (i.e., the clusters) have grown in
mass  (see also Brown  \etal 2008).   Finally, Kang  \& van  den Bosch
(2008) argue  in favor of  the creation of  stellar halos in  order to
prevent central galaxies from accreting  too many blue and/or gas rich
satellites, which would cause central galaxies to be too blue.

 Recently Purcell  et al.   (2007)  used the  analytic model  for
  subhalo infall and evolution  of Zentner \etal (2005), combined with
  empirical  constraints on  the  stellar mass  fractions of  accreted
  subhalos,  to predict  the diffuse  stellar mass  fractions  of dark
  matter halos.   They predict that the average  stellar mass fraction
  in diffuse, intrahalo light increases strongly from $\sim 0.005$ for
  small galaxy  halos ($\sim 10^{11}  \msunh$) to $\sim 0.2$  for poor
  groups  ($\sim 10^{13}  \msunh$), after  which the  trend  with mass
  flattens considerably.   A comparison  with the right-hand  panel of
  Fig.~\ref{fig:MM} shows that this is in remarkable agreement with our
  predictions, which  have been  deduced using a  completely different
  approach.  In  addition to this, Purcell \etal  (2007) also compared
  their model predictions to  the observational data of Gonzales \etal
  (2007), similar to our comparison in Fig. \ref{fig:ICS}, and reached
  a  similar conclusion  as ours,  namely  that the  vast majority  of
  subhalo  stars must  be  deposited into  a  diffuse halo  component.
  Similar results  have also been  obtained by Henriques  \etal (2008)
  using a  semi-analytical model for galaxy formation  that includes a
  simple treatment  for the disruption  of satellite galaxies.   It is
  reassuring that different approaches  yield results that are in such
  good agreement.

\begin{figure}
\plotone{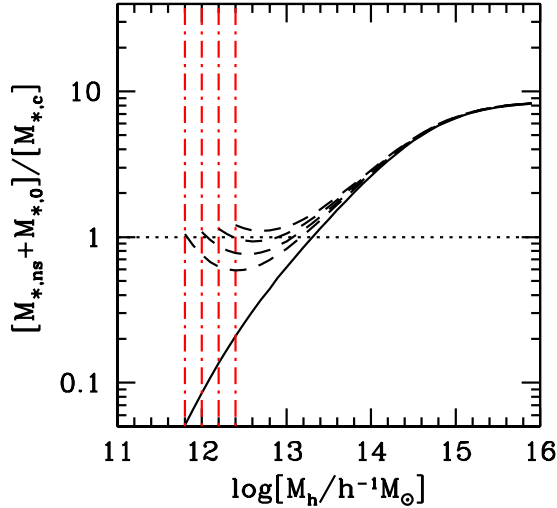}
\caption{ The  ratios of the  stripped or disrupted stellar  mass from
  satellite  galaxies  together  with  the  stellar  mass  of  central
  galaxies at the quench point, $M_{\rm \ast, ns}+M_{\rm \ast, 0}$, to
  that  of  the  central  galaxy,  $M_{\rm \ast,  c}$.  See  text  for
  details. }
\label{fig:quench}
\end{figure}

\subsection{A Lower Limit on the Halo Mass for Quenching Star Formation}
  
In hierarchical models of structure formation, massive halos are built
up by  the mergers  of smaller  ones.  Over the  years, it  has become
clear  that  a  successful  model  for  galaxy  formation  in  such  a
hierarchical  framework  requires a  mechanism  that  can quench  star
formation in  massive galaxies. Currently, the  most favored quenching
mechanism is feedback from an  active galactic nucleus (AGN), which is
often assumed to operate above  a given halo mass, $\Mq$ (e.g.  Croton
\etal 2006; Cattaneo  \etal 2006, and references therein).  We can use
the results  presented above  to put a  lower limit on  this quenching
mass $\Mq$, as follows.

In general, the stars in a  central galaxy may come from two different
channels:  (i)  {\it  in  situ}  formation, and  (ii)  accretion  from
satellite  galaxies   (those  that  are  part   of  the  non-surviving
component).   Consider a  model in  which star  formation  is abruptly
quenched in halos with $\Mh > \Mq$.  In such quenched halos, the total
amount of  stars from channel  (i) is at  most equal to  $M_{\rm \ast,
  0}$, the stellar mass contained in the central galaxy of a halo with
mass $\Mq$.  It may be smaller than $M_{\rm \ast, 0}$, because part of
this mass may be due to accretion rather than {\it in situ} formation.
The total amount  of stars from channel (ii) is  at most $M_{\rm \ast,
  ns}$, because part of the disrupted  mass may end up in the ICS.  In
Fig.  \ref{fig:quench}, we show the  ratio $(M_{\rm \ast, ns} + M_{\rm
  \ast, 0})/ M_{\rm \ast, c}$,  as a function of halo mass.  Different
lines  show the  results for  different values  of $\Mq$  [$=0$ (solid
  line)  , $10^{11.8}$, $10^{12.0}$,  $10^{12.2}$, $10^{12.4}\msunh$].
As one  can see, if  $\Mq\la 10^{12.4}\msunh$, the total  stellar mass
from  the two  channels,  $M_{\rm \ast,  ns}  + M_{\rm  \ast, 0}$,  is
insufficient to account for the stellar masses of the central galaxies
in  halos  with  $\Mh  \sim 10^{12.5}\msunh$.   However,  for  $\Mq\ga
10^{12.4}\msunh$  the ratio  $(M_{\rm \ast,  ns} +  M_{\rm  \ast, 0})/
M_{\rm \ast,  c} >  1$ for all  halo masses.   This implies that  if a
quenching  mass $\Mq$  exists,  it must  be  larger than  or equal  to
$10^{12.4} \msunh$.

\section{Discussion and Summary}
\label{sec_summary}

Using the  conditional stellar  mass functions for  satellite galaxies
obtained by Yang et al.  (2008b) and the sub-halo mass functions given
by Giocoli  et al.  (2008),  we study the connection  between subhalos
and  satellite  galaxies.  Assuming  that  at  the  time of  accretion
satellite galaxies are associated  with subhalos according to the same
stellar mass - halo mass relation as present-day central galaxies with
halos, we predict the stellar  mass function of satellite galaxies and
compare our model predictions  with observations. Our main results can
be summarized as follows:
\begin{itemize}
\item Assuming  that the stellar  masses of satellite galaxies  do not
  evolve after they are accreted into the host halos, we find that the
  model over-predicts the population of satellite galaxies, especially
  in  low mass  halos.  One  solution,  albeit unlikely,  is that  the
  stellar mass fractions of haloes  at higher redshifts are lower than
  at  the present.   The other,  which  is considered  in the  present
  paper,  is that a  significant fraction  of satellite  galaxies have
  been stripped of their stars or even totally disrupted.
\item Assuming that the amount  of disruption of satellite galaxies is
  a function of the ratio between  the mass of the subhalo and that of
  the host halo, we find  that the surviving fraction can be described
  by $f(x = \log(m_s/M_h))  = -0.041 -0.820x -0.422x^2 -0.078x^3$. The
  decrease  of $f$ with  increasing $m_s/M_h$  is consistent  with the
  idea that dynamical friction  brings subhalos and satellite galaxies
  to the  inner part of the  host halo where  stripping and disruption
  are more efficient.
\item The  majority of the  stars stripped from satellites  in massive
  halos  are predicted  to  end  up as  intra-cluster  stars, and  the
  predicted amounts  of the intra-cluster  component as a  function of
  the   velocity  dispersion   of   galaxy  system   match  well   the
  observational  results obtained  by  Gonzalez et  al.  (2007).   The
  fraction of the mass in the  stripped stars to that in the surviving
  central and satellite galaxies is predicted to be the highest ($\sim
  40\%$ of the total) in  halos with masses $\sim 10^{14} \msunh$.  If
  all  these stripped  stars  end up  in  the intra-cluster  component
  (Max),  or maximum  of them  are  accreted into  the central  galaxy
  (Min), then we can predict that  a maximum $\sim 19\%$ and a minimum
  $\sim 5\%$ of the total stars  in the whole universe are in terms of
  the diffused intra-cluster component.
\item  Stars stripped from  satellite galaxies  are not  sufficient to
  build   up  the  central   galaxies  in   halos  with   masses  $\la
  10^{12.5}\msunh$, and  so star formation  should not be  quenched in
  halos with masses up to at least $\sim 10^{12.4}\msunh$.
\end{itemize}

It should be pointed out once  again that our results are based on the
assumption  that subhalos  have  the  same stellar  mass  - halo  mass
relation as the  host halos at redshift $z=0$. If  the total amount of
stars that can  form in a halo at higher redshift  is larger than that
in a halo of the same  mass at lower redshift (e.g., Cooray 2005), the
predicted disrupted  fraction of  satellite galaxies would  be larger,
and so would be the predicted amount of ICS. On the other hand, if the
total amount  of stars that can  form in high-$z$ halos  is lower than
that in  its $z=0$ counterparts,  the predicted ICS fraction  would be
lower.  It  is interesting that  the observed ICS fraction  is matched
well with the assumption that the  total amount of stars that can form
in a halo of a given mass is independent of redshift.  This assumption
is also  consistent with the results  obtained by Wang  et al.  (2006)
based  on a  semi-analytical model  of structure  formation.  However,
current  semi-analytical   models  of  galaxy   formation  ignore  the
existence  of the  ICS component,  so that  the stars  that  formed in
subhalos  are assumed  either to  remain in  satellite galaxies  or to
merge into central galaxies.  Consequently, such models either predict
too  high  a  mass  for  the  central galaxies  in  massive  halos  or
over-predict the number of satellite galaxies in group-sized halos.
 
As  described above, the  mass fraction  of stars  in the  diffuse ICS
component is much  larger than that in the  central galaxies for halos
with masses $\ga 10^{14}\msunh$, and  even in the whole universe, this
stellar  mass  component is  a  significant  fraction  of $\sim  19\%$
(upper-limit)  or $\sim  5\%$  (lower-limit). The  implication of  the
existence of such a stellar  component for galaxy formation has yet to
be explored. Indeed, if the  total amount of stars in groups, clusters
and  in the  universe  is larger  than  that implied  by the  observed
stellar mass  function of galaxies, and if  semi-analytical models and
numerical simulations do not resolve  the ICS component, then it would
be incorrect  to use the observed stellar  mass/luminosity function of
galaxies  to   constrain  star  formation   efficiency  and  feedback.
Furthermore, the fraction  of the ICS component in  halos of different
masses may  also convey important  information about the  evolution of
galaxies in different environments.  Clearly, more theoretical work is
required in order to make full  use of the information provided by the
ICS component.

                                                                           
\acknowledgments We  are grateful to the anonymous  referee for useful
and insightful comments that helped to improve the presentation.  This
work is supported  by the {\it One Hundred  Talents} project, Shanghai
Pujiang Program (No.  07pj14102), 973 Program (No.  2007CB815402), the
CAS Knowledge Innovation Program  (Grant No.  KJCX2-YW-T05) and grants
from  NSFC (Nos.  10533030,  10673023, 10821302).   HJM would  like to
acknowledge the  support of NSF AST-0607535, NASA  AISR-126270 and NSF
IIS-0611948.


\clearpage

\end{document}